\documentstyle[aps,prl,multicol]{revtex}
\begin{document}
\author{X. X. Yi, G. R. Jin, D. L. Zhou}
\address{ Institute of Theoretical Physics, Academia Sinica, P.O.Box 2735, Beijing 100080, China}
\title{Creating Bell states and decoherence effects in quantum dots
system } \maketitle
\begin{abstract}
We show how to improve the efficiency for preparing Bell states in
coupled two quantum dots system. A measurement to the state of
driven quantum laser field leads to wave function collapse. This
results in highly efficiency  preparation of Bell states. The
effect of decoherence on the efficiency of generating Bell states
is also discussed in this paper. The results show that the
decoherence  does not affect the relative weight of $|00\rangle$
and $|11\rangle$ in the output state, but the efficiency of
finding Bell states.

{\bf PACS number(s): 03.67.-a,71.10.Li,71.35.-y}
\end{abstract}
\vspace{4mm}
\begin{multicols}{2}[]

The EPR paradox was proposed in the earliest days of quantum
mechanics[1]. Since then quantum entanglement became a interesting
subject, which lies at the heart of the profound difference
between quantum and classical physics. The experimental
realization of quantum entangled state may be traced back to the
mid 1960's, when entangled photon pairs created from cascade
emission[2]. The primary motivation for creating entangled states
was to test Bell's inequality[3], which was derived by using the
hidden variables theory. In recent years, research involving
entangled photon pairs and other entangled pairs, such as atom's
has proceeded away from the foundations of quantum mechanics and
to applications in quantum cryptography[4] and quantum
teleportation[5,6].

Creating entangled states is the first step toward studying any
effects related to  entanglement. At the present time, the most
widely used technique in experiments for generating entangled
pairs is spontaneous parametric down-conversion in a nonlinear
crystal[7]. The entangled photon pairs, however, are normally
difficult to store in contrast to  massive particles. To overcome
this difficulty, several proposals including correlated
atom-photon pairs, non-classical multi-atom states[10-13] and
entangled atom-pair from atomic Bose-Einstein condensate[14-19]
are proposed. Semiconductor quantum dots(QDs) have their own
advantages as a candidate of the basic building blocks of
solid-state based quantum logic devices, due to the existence of
an industrial base for semiconductor processing and the ease to
integration with existing device[20-22]. Experimental realization
of optically induced entanglement of excitons in a single quantum
dot[23] and theoretical study on coupled quantum dots[24] was
reported most recently. In the reports, a classical laser field is
applied to create the electron-hole pair in the dot(s). In this
paper, we generalize the proposal[24] to the case of quantum laser
field, instead of the classical laser filed. The scheme under our
consideration has a potential advantage that we obtain highly Bell
states of excitons through measuring the state of the quantum
laser field. Besides, the decoherence caused by acoustic
phonon-exciton interaction is also studied in this paper.

To begin with, we consider the Hamiltonian[24]
\begin{equation}
H=eJ_z+W(J^2-J_z^2)+\xi(t)J_++\xi^*(t)J_-,
\end{equation}
where
\begin{eqnarray}
J_+=\sum_{i=1}^Nc_i^{\dagger}h_i^{\dagger}, \ \
J_-=(J_+)^{\dagger},\nonumber\\ J_z=\frac 1 2
\sum_{i=1}^N\{c^{\dagger}_ic_i-h_ih^{\dagger}_i\},
\end{eqnarray}
$c^{\dagger}_i(h^{\dagger}_i)$ is the electron (hole) creation
operator in the ith quantum dot, $e$ is the QD band gap, $W$
stands for the interdot interaction,  and $\xi(t)$ denotes the
laser pulse times the strength of the coupling of  the laser field
to the excitons. This Hamiltonian governs the time evolution of
the excitons in $N$ identical quantum dots system with interdot
transfer. The $J_i$ operator obey the usual angular momentum
commutation relations $[J_z, J_{\pm}]=\pm J_{\pm}, [J_+,
J_-]=2J_z.$ We consider a laser field  $\xi(t)=a A e^{-i\omega
t}+a^{\dagger}A^*e^{i\omega t}$ with electron-photon coupling and
the electric field strength $A$, the frequency $\omega$, and the
photon creation (annihilation) operator $a^{\dagger}$ ($a$).  For
a coupling two quantum dots system, we introduce notions
$|0\rangle=|J=1, M=-1\rangle$, $|1\rangle=|J=1, M=0\rangle$,
$|2\rangle=|J=1, M=1\rangle$ to describe the vacuum state, the
single exciton state and the biexciton state, respectively. In
terms of $|0\rangle$, $|1\rangle$ and $|2\rangle$, the Hamiltonian
(1) can be re-expressed as
\begin{equation}
H_e=\sum_{i=0}^2 E_i|i\rangle\langle i|+ \sqrt{2}A a|1\rangle
\langle 0| +\sqrt{2}A a|1\rangle \langle 2|+H.c,
\end{equation}
where $E_0=W-e+0.5\omega, E_1=2W-0.5\omega, E_2=W+e+0.5\omega.$We
now show that this Hamiltonian leads to the generation of Bell
states from suitably initialized states. In an invariant subspace
spanned by $|1,n\rangle=|1\rangle\otimes|n\rangle$, $|0,n+1\rangle
=|0\rangle\otimes|n+1\rangle$ and
$|2,n+1\rangle=|2\rangle\otimes|n+1\rangle.$ The Hamiltonian takes
the following  form
\begin{equation}
H_e=\left( \matrix{ E_1 & \sqrt{2(n+1)}A & \sqrt{2(n+1)}A\cr
\sqrt{2(n+1)}A^* &E_0 &0\cr \sqrt{2(n+1)}A^* & 0 & E_2} \right ),
\end{equation}
where $|n\rangle$ is Fock state of the laser field. For
explicitness, we study here the case of $E_0\sim E_2$, and choose
$A=A^*$. Defining $\Omega_1=\sqrt{2(n+1)} A,$  and
$\Omega^2=8\Omega_1^2+(E_1-E_0)^2$, the Hamiltonian can be
rewritten as

\begin{eqnarray}
H_e&=&E_0+\nonumber\\ &\ &\Omega\left( \matrix{ \cos\theta & \frac
{\sqrt{2}}{4} \sin\theta & \frac{\sqrt{2}}{4} \sin\theta\cr
\frac{\sqrt{2}}{4} \sin\theta &0 &0\cr \frac{\sqrt{2}}{4}
\sin\theta & 0 & 0} \right ), \nonumber\\
\end{eqnarray}
with $ \tanh\theta=\frac{2\sqrt{2}\Omega_1}{E_1-E_0}.$ Then, the
eigenvalues and the corresponding eigenstates are
\begin{eqnarray}
E_d(n)&=&E_0,\nonumber\\ E_{\pm}(n)&=&\Omega\frac{\cos\theta\pm
1}{2}+E_0,
\end{eqnarray}
and
\begin{eqnarray}
|E_d\rangle&=&\frac{\sqrt{2}}{2}(|0\rangle-|2\rangle)\otimes|n+1\rangle\nonumber\\
&\equiv& B_{d,0}|0,n+1\rangle+B_{d,2}|2,n+1\rangle,\nonumber\\
|E_+\rangle&=&\frac{\sqrt{2}}{2}\sin\frac{\theta}{2}(|0\rangle+|2\rangle)\otimes|n+1\rangle
+\cos\frac {\theta}{2}|1,n\rangle\nonumber\\ &\equiv&
(B_{+,0}|0\rangle+B_{+,2}|2\rangle)\otimes|n+1\rangle+B_{+,1}|1,n\rangle,\nonumber\\
|E_-\rangle&=&-\frac{\sqrt{2}}{2}\cos\frac{\theta}{2}(|0\rangle+|2\rangle)\otimes|n+1\rangle
+\sin\frac {\theta}{2}|1,n\rangle\nonumber\\ &\equiv&
(B_{-,0}|0\rangle+B_{-,2}|2\rangle)\otimes|n+1\rangle+B_{-,1}|1,n\rangle,
\end{eqnarray}
respectively. The above solution shows that the system of the
exciton with $J=1$, and $\Delta=0$ has a dark state $|E_d\rangle$,
which decouples with the state $|1\rangle$. In the process of
adiabatic evolution the probability of transition from
$|E_d\rangle$ to $|1,n+1\rangle$ is zero when the system is
initially in the dark state $|E_d\rangle$.

Let us consider an initial state $|0,n+1\rangle$, i.e., the
excitons are in their vacuum state,  while the laser field is in a
Fock state $|n+1\rangle$ initially. At time $t$, the wave function
of system that consists of the excitons and the laser field is
\begin{eqnarray}
|\psi(t)\rangle&=&[\frac 1 2 \sin^2\frac{\theta}{2}
e^{-iE_+(n)t}(|0\rangle+|2\rangle)\nonumber\\ &+& \frac 1 2
\cos^2\frac{\theta}{2}(|0\rangle+|2\rangle)
e^{-iE_-(n)t}\nonumber\\ &+& \frac 1 2
e^{-iE_d(n)t}(|0\rangle-|2\rangle)]\otimes|n+1\rangle\nonumber\\
&+&\frac {\sqrt{2}}{2}
[\sin\frac{\theta}{2}\cos\frac{\theta}{2}e^{-iE_+(n)t}\nonumber\\
&-& \cos\frac{\theta}{2}\sin\frac{\theta}{2}
e^{-iE_-(n)t}]|1,n\rangle.
\end{eqnarray}
The probability of finding one exciton in the two coupled
simeconductor quantum dots  is
\begin{equation}
P_1(t)=\sum_nP(n+1)\sin^2\frac{\theta}{2}\cos^2\frac{\theta}{2}\sin^2\frac{\Omega
t}{2},
\end{equation}
where we choose  $|\alpha\rangle=\sum_nP(n)|n\rangle$ and
$|0\rangle$ as the initial states of the laser field and the
excitons, respectively.
 The probability $P_1(t)$ versus time $t$ is
illustrated in figure 1. As seen from figure 1, there are
evidences of collapses and revivals in the coupled quantum dots
system. The collapse time obviously depend on the strength of the
driven laser field. The stronger the laser field, the shorter the
collapse time (from fig.1,a to fig.1,b). Despite the fact that
there is no experimental study on the collapses and revivals in
this system. The observation of excitonic Rabi oscillations in
semiconductor quantum well[25] lead us to believe that we are not
too far away from the experimental observation of the existence of
this effects.
 Now we take a measure
to the laser field, if we observe that the laser field is in its
initial state $|n+1\rangle$, then the excitons are in
state(unnormalized)
\begin{eqnarray}
|\phi(t)\rangle&=&\frac 1 2 \sin^2\frac{\theta}{2}
e^{-iE_+(n)t}(|0\rangle+|2\rangle)\nonumber\\ &+& \frac 1 2
\cos^2\frac{\theta}{2}(|0\rangle+|2\rangle)
e^{-iE_-(n)t}\nonumber\\ &+& \frac 1 2
e^{-iE_d(n)t}(|0\rangle-|2\rangle)
\end{eqnarray}
certainly, this is a superposition of Bell state $$
\frac{1}{\sqrt{2}}(|00\rangle+|11\rangle),\mbox {and}
\frac{1}{\sqrt{2}}(|00\rangle-|11\rangle).$$ Figure 2 shows the
ratio of $P_+(t)$ to $P_-(t)$ versus time $t$ for the initial
condition $|\psi(0)\rangle=|0,n+1\rangle$. Where $P_{\pm}(t)$  is
the probability for finding the Bell state
$\frac{1}{\sqrt{2}}(|00\rangle\pm|11\rangle)$. As seen from figure
2, selective pulses of length $T$ can be used to create wanted
Bell states $\frac{1}{\sqrt{2}}(|00\rangle\pm|11\rangle)$ in the
systems of two coupled QDs. For example, if the length of the
laser field satisfies $\cos E_+(n) T/\cos
E_-(n)T=ctanh^2\theta/2$, the output state is exactly in Bell
state $\frac{\sqrt{2}}{2}(|0\rangle-|2\rangle)$.  Figure 2-b is
the same as figure 2-a, but with different initial photon number
$n$, while energy $W$ is kept fixed in  figure 2. The results
indicate that the time $T$ is increased with decreasing the photon
number $n$ (from figure 2-a to figure 2-b) .

It is well known that the decoherence is the most enemy to prepare
entangled state and keep a state entangled in the quantum
information processing. In what follows, we analyze  the
reliability of the preparation of Bell states when decoherence is
take into account, the decoherence of excitons in semiconductor
quantum dots mainly results from the electron-accoustic phonon
interactions, this process is governed by the Hamiltonian[24]
\begin{equation}
{\cal H}=H+\sum_k\omega_kb_k^{\dagger}b_k+\sum_k
g_kJ_z(b_k^{\dagger}+b_k),
\end{equation}
where $H$ is given by eq.(2), and $b_k^{\dagger}(b_k)$ stands for
the creation (annihilation) operator of the acoustic phonon with
wavevector $k$. By the general procedure, we can deduce a master
equation for the reduced density operator under Markov
approximation
\begin{equation}
i\frac{\partial}{\partial t}\rho=[H,\rho]-i\Gamma[J_z[J_z,\rho]],
\end{equation}
where $\Gamma$ is the decoherence rate which depends on the mode
distribution of phonons as well as the cut-off frequency.
Furthermore, we assume that $\Gamma/A<<1$, i.e., the decoherence
rate $\Gamma$ is much smaller than the coupling between the
excitons and the laser field, this assumption allows us to solve
the master equation by small-loss expantions[26]. To begin with,
we write the density operator that is a solution to the master
equation in powers of $\Gamma$
\begin{equation}
\rho(t,\Gamma)=\rho(t,0)+\rho_1 \Gamma+\frac 1 2
\rho_2\Gamma^2+...,
\end{equation}
where $\rho_1=\frac{\partial \rho}{\partial\gamma}|_{\Gamma=0},$
$\rho_2=\frac{\partial ^2\rho}{\partial \Gamma^2}|_{\Gamma=0}$.
Substituting this expantions into the master equation, we obtain
the following set of equations
\begin{eqnarray}
i\dot{\rho}(t,0)&=&[H,\rho(t,0)],\\ i\dot{\rho}_1&=&
[H,\rho_1]-i[J_z[J_z,\rho(t,0)]],\\ i\dot{\rho}_2&=&
[H,\rho_2]-i[J_z[J_z,\rho_1]],
\end{eqnarray}
In general, we can solve eq.(14) exactly for an given initial
condition, which gives the zeroth order solution $\rho(t,0)$.
Substituting the zeroth order equation into eq.(15), $\rho_1$ can
be calculated. Following this procedure, successive terms of the
expantion could be worked out, though the calculation is
complicated. The average value of a given operator, say $B$, could
be calculated through
\begin{equation}
\langle B\rangle=Tr (\rho B)=Tr(\rho(t,0)B)+\Gamma Tr(\rho_1
B)+...,
\end{equation}
For a initial state $|0,n+1\rangle$, it is obvious from eq.(14)
that
\begin{equation}
\rho(0,t)=|\psi(t)\rangle\langle \psi(t)|,
\end{equation}
where $|\psi(t)\rangle$ is given by eq.(8). A straightforward
calculation shows that
\begin{eqnarray}
\rho_1^{a,b}&\equiv& \langle E_a|\rho_1|E_b\rangle,
a,b=d,+,-\nonumber\\ &=&-\int_0^t d\tau
f_{a,b}e^{i(E_a-E_b)(t-\tau)},
\end{eqnarray}
where
$$f_{a,b}=4B_{a,0}B_{b,0}\rho^{0,2}(t,0)+4B_{a,2}B_{b,2}\rho^{2,0}(t,0),
$$ $$\rho^{0,2}(t,0)=\langle
0,n+1|\rho(t,0)|2,n+1\rangle=B_{0,0}B_{2,0}e^{i(E_0-E_2)t}, $$
$$\rho^{0,2}(t,0)=[\rho^{2,0}(t,0)]^*,$$ Equations(19)and (13)
together give
\begin{equation}
\rho^{a,b}(t,\Gamma)=\rho^{a,b}(t,0)-\Gamma\int_0^t d\tau
f_{a,b}e^{i(E_a-E_b)(t-\tau)},
\end{equation}
it is easy to show that the output state after measuring the laser
field state $|n+1\rangle$ is
\begin{eqnarray}
\rho_e&=&Tr_f(|n+1\rangle\langle n+1|\rho(t,\Gamma))\nonumber\\
&=& |\phi(t)\rangle\langle
\phi(t)|-\Gamma\sum_{a,b=+,-,d}\rho_1^{a,b}(t)(B_{a,0}|0\rangle+B_{a,2}|2\rangle)\nonumber\\
&\cdot &(\langle 0|B^*_{b,0}+\langle 2|B^*_{b,2}),
\end{eqnarray}
The first term is just the output state (10) in the situation
without decoherence, while the last are represent the effect of
decoherence on the generating Bell states. In the case of exact
resonance, $f_{a,b}$ is time independent, hence $\rho_1^{a,b}$ for
$a=b$ is proportional to $t$. For $a\neq b$,
$$\rho_1^{a,b}=\frac{if_{a,b}}{E_a-E_b}(1-e^{-i(E_a-E_b)t}).$$
Noticing the second term in eq.(21) is also a superposition of
Bell states, we come to a conclusion that the decoherence does not
influence the relative weight of $|00\rangle$ and $|11\rangle$,
but diminish the efficiency of finding Bell state. For instance,
the probability of finding Bell state
$\frac{1}{\sqrt{2}}(|00\rangle-|11\rangle)$ is
$P_-=\frac{1}{2}-\Gamma f_{dd} t$, which decrease proportionally
with time.

In summary, we generalized the proposal proposed in [24] to the
case of quantum laser field, the advantages of this generalized
proposal are that we can select the output state through measuring
the laser field state. This provide with us a method to generate
Bell states with highly efficiency. Decoherence caused by acoustic
phonon scattering diminish the efficiency of preparing Bell
states. Up to the first order of $\Gamma$, the decrease in
probability of finding Bell state
$\frac{\sqrt{2}}{2}(|0\rangle-|2\rangle$ is proportional to time
$t$. Generally speaking, the dependence of the decrease on time is
complicated, although we only consider the first order on
$\Gamma$.\\
 {\bf \large ACKNOWLEDGEMENT:}\\ This work is supported
by the Chinese postdoctoral Fund, and NNSF of China.\\

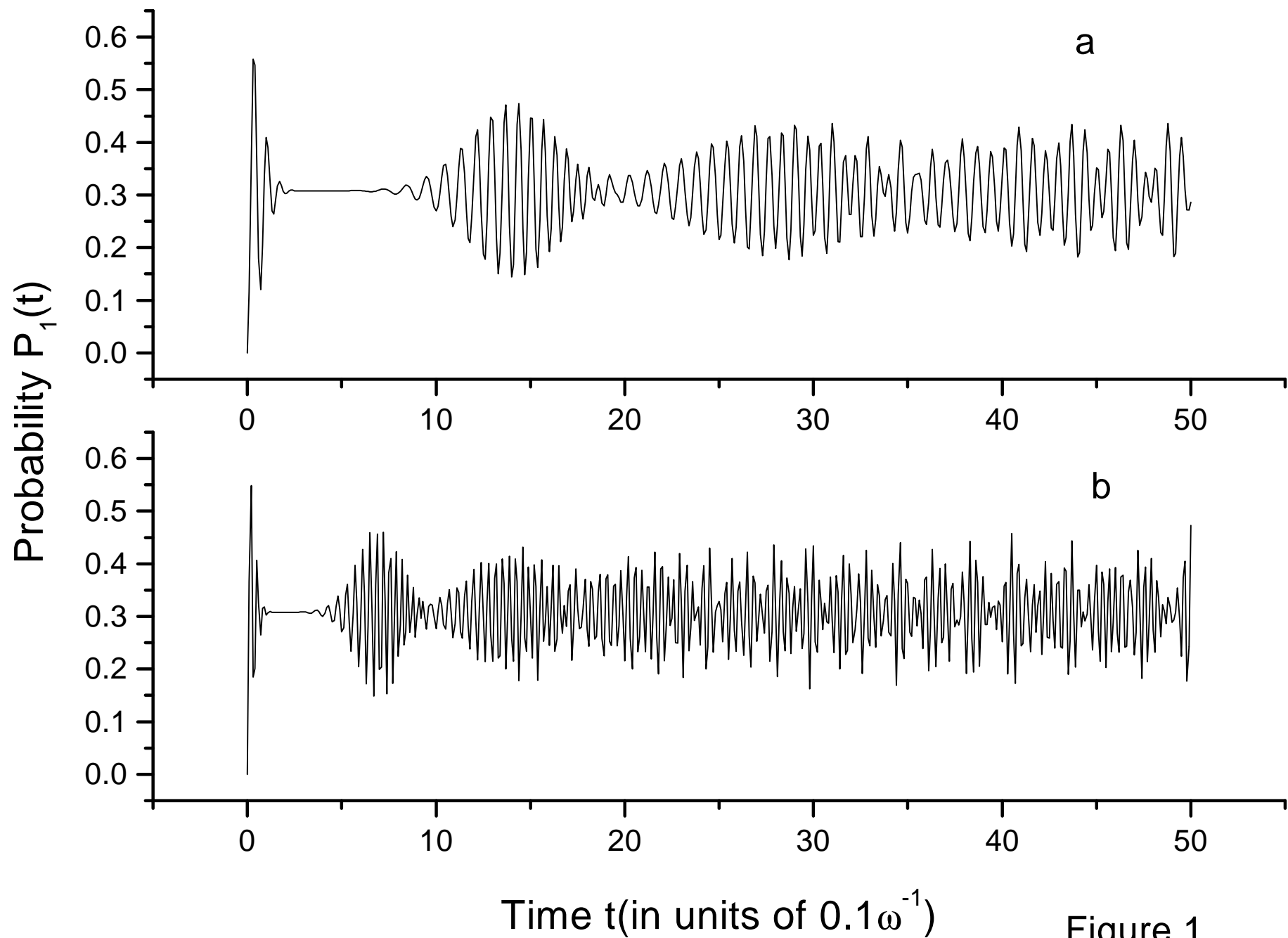
{\bf Figure captions}\\ {\bf Fig.1}: Population of the one exciton
state in two coupled QDs, as a function of time. The parameter
chosen are a:$\alpha=5, \omega=10^{15}, W=0.1\omega, A=0.4W$, b:
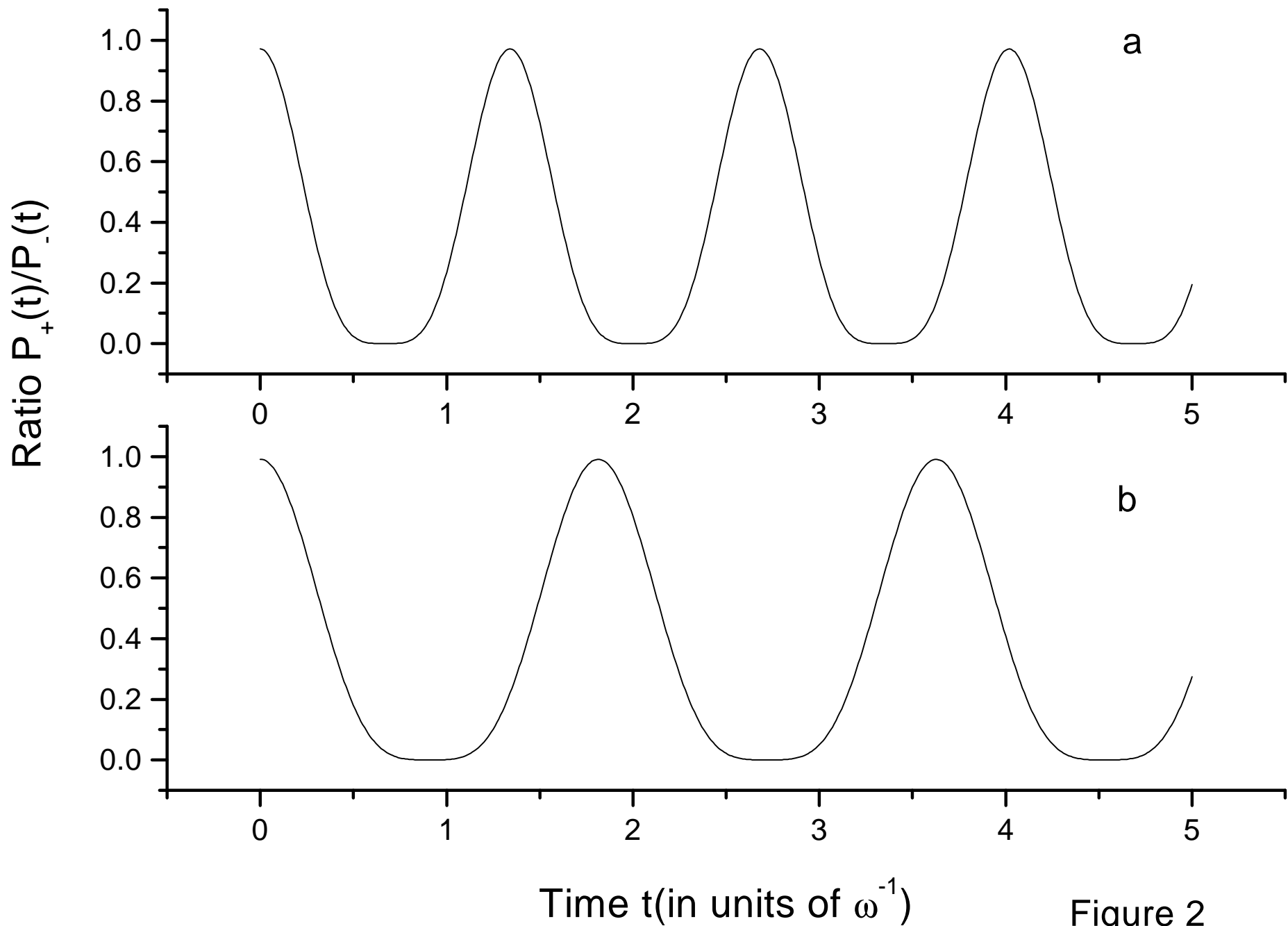
are the same as those in $a$, but with $A=0.8W$.\\ {\bf Fig.2}:
The ratio of the pupulation in Bell state
$\frac{\sqrt{2}}{2}(|00\rangle+|11\rangle$ to that in
$\frac{\sqrt{2}}{2}(|00\rangle-|11\rangle$. We choose a:$n=10$,
b:$n=5$. Ther other parameters are the same as those in figure 1.
\end{multicols}
\end{document}